\documentclass[12pt]{article}
\def \be {\begin{equation}}
\def \ee {\end{equation}}
\def \bea {\begin{eqnarray}}
\def \eea {\end{eqnarray}}
\def \nn {\nonumber}

\def \la {\langle}
\def \ra {\rangle}

\def \del {\partial}
\def \dels {\partial\kern-.5em / \kern.5em}
\def \As {{A\kern-.5em / \kern.5em}}
\def \Ds {D\kern-.7em / \kern.5em}

\def \a {\alpha}

\def \dag {\dagger}

\def \d {\delta}

\def \m {\mu}
\def \n {\nu}

\def \lam {\lambda}

\def \s {\sigma}

\def \om {\omega}

\def \th {\theta}
\def \Th {\Theta}

\def \II {I\hspace{-.1em}I\hspace{.1em}}

\def \IIB {\mbox{\II B\hspace{.2em}}}
\def \bm{\bibitem}

\newcommand{\ket}[1]{|#1 \rangle}
\newcommand{\eq}[1]{(\ref{#1})}

\def \P {\vec{P}}
\def \rra {\ra\hspace{-.2em}\ra\hspace{.0em}}
\def \lla {\la\hspace{-.2em}\la\hspace{.0em}}

\begin{document}
\begin{titlepage}
%\catcode`\@=11
%\catcode`\@=12
%\twocolumn[\hsize\textwidth\columnwidth\hsize\csname%
%@twocolumnfalse\endcsname

%\draft
\begin{center}
\hfill hep-th/0205218\\
\vskip .5in

\textbf{\Large Cubic String Field Theory in pp-wave Background
and Background Independent Moyal Structure}

\vskip .5in
{\large Chong-Sun Chu$^1$, Pei-Ming Ho$^2$, Feng-Li Lin$^3$}
\vskip 15pt

{\small \em $^1$
Department of Mathematical Sciences,
University of Durham, Durham, DH1 3LE, UK
}\\
{\small \em $^2$Department of Physics,
National Taiwan University,
Taipei, Taiwan, R.O.C.}\\
{\small \em $^3$Department of Physics,
Tamkang University, Tamsui, Taiwan, R.O.C.}

\vskip .2in \sffamily{chong-sun.chu@durham.ac.uk}, \sffamily{
pmho@phys.ntu.edu.tw}, \sffamily{ linfl@mail.tku.edu.tw}
\vspace{60pt}
%\maketitle
\end{center}
\begin{abstract}

We study Witten open string field theory
in the pp-wave background in the tensionless limit,
and construct the $N$-string vertex
in the basis which diagonalizes the
string perturbative spectrum.
We found that the Witten $*$-product
can be viewed as infinite copies of the
Moyal product
with the same noncommutativity parameter $\th=2$.
Moreover, we show that this Moyal structure
is universal in the sense that,
written in the string bit basis, Witten's
$*$-product for any background
can always be given in terms of the above-mentioned Moyal structure.
We identify some projective operators in this algebra
that we argue to correspond to D-branes of the theory.

\end{abstract}
%\pacs{PACS numbers: 11.25.-w, 11.25.Mj, 11.25.Sq}%]
\end{titlepage}
%\begin{narrowtext}
\setcounter{footnote}{0}

\section{Introduction} \label{basic}

Recently, various aspects of string theory
in pp-wave background are being studied with immense interests.
The is partially due to the remarkable proposal
\cite{BMN}
which states that a sector of the SYM operators
with large $R$-charge is dual to the \IIB string
theory on a pp-wave background consisting of the metric
\be \label{pp-metric}
ds^2 = -\mu^2 x^i x^i (dx^+)^2 -4 dx^+ dx^- +
\sum_{i=1}^{D-2} dx^i dx^i,
\ee
with $D-2=8$ transverse directions and a homogeneous RR 5-form flux.

The study of pp-wave background as an
exact solution
of string theory has a long history
(see e.g. \cite{pp1}-\cite{pp4} and \cite{pp-rev} for review and
references therein). It was realised only recently
that the pp-wave background is maximally
supersymmetric \cite{bfhp} in the presence of a certain RR flux. Moreover the
string model is exactly solvable \cite{metsaev,t1},
just as in the NS-NS case.
In this paper, we will concentrate on the simplest form of the pp-wave
metric \eq{pp-metric}.
The metric can be supported by different combinations of
NS or RR flux and leads to different exactly solvable string models
\cite{metsaev,t1,KK}.
For example, in the maximally
supersymmetric \IIB
pp-wave background \cite{bfhp}, an RR flux
$H_{+1234}= H_{+5678} = \mu$ is turned on.
In the Nappi-Witten background (D=4) \cite{NW} and its higher dimensional
generalization \cite{KM}, the target space is the extended Heisenberg group
$H_n$ whose dimension is $D=2n+2$; and the background is supported by an NS
flux $H_{+12} = \cdots = H_{+ (2n+1)(2n+2)} = \mu$. These
backgrounds are distinguished because they are exactly solvable.
Here ``exactly solvable'' means that
the model can be written in terms of free
oscillators in the light cone gauge.

One of the main motivations in studying string field theory is the
hope of a better understanding of the nonperturbative
phenomena in string theory.
Remarkable progress has been
made recently in
understanding D-branes and tachyon condensation
\cite{VSFT} from the
viewpoint of open string field theory \cite{Witten}. In Witten's
formulation, the open string $*$-product of fields $\Psi_1, \Psi_2$ is
defined by identifying the right half of the string of $\Psi_1$ with
the left half of the string of $\Psi_2$. The idea is made concrete
\cite{GJ1} by using conformal theory techniques to
provide a Fock space description of the Witten $*$-product.
Recently, the work of Bars \cite{Bars} and Douglas, Liu, Moore and
Zwiebach \cite{DLMZ} has led to
remarkable progress in identifying the Witten $*$-product as a
continuous Moyal product. This new formulation
may have the advantage in understanding the nonperturbative
symmetries of string field theory based upon the progress of
noncommutative field theory.

In this paper, we study the Witten open string
field theory in the pp-wave background.
In particular we would like to construct
the Witten $*$-product as a first step to
understand the algebraic structure of the string field theory.
To avoid the complications due to picture changing operators
in the supersymmetric case \cite{Witten2}
(see for example \cite{Berkovits} for more discussions),
we will work in the bosonic case with
the Nappi-Witten type background $H_n$.
Since the string model is simplified and can be written in
terms of free oscillators only in the light cone gauge,
one can try to construct the covariant open string $*$-product
in certain limiting situation.
Inspired by the considerations in \cite{T-limits}
(where the motivation was to consider specific limit,
particularly the tensionless limit, such that
the $AdS_5 \times S^5$ string model
gets simplified and further understanding of
the $AdS$/CFT correspondence can be achieved),
we consider the tensionless limit of the bosonic string
in the Nappi-Witten type pp-wave background.
It has also been argued \cite{Lind} that
the tensionless limit should provide
the proper starting point for investigating
the high energy limit of string dynamics.

Witten's
$*$-product is defined by the overlap conditions
and is
% p3 an invariant concept.
a background-independent concept.
However the explicit representation of it,
and hence the associated Moyal product,
is basis dependent and background dependent.
For the purpose of perturbative studies,
it is convenient to use a basis which
diagonalizes the Hamiltonian.
For example, in the flat background,
% p3
the
Witten $*$-product was constructed \cite{GJ1} in terms of
the {\it harmonic oscillator basis} which
diagonalizes the Hamiltonian. In the tensionless
limit, the string worldsheet is described most naturally in terms of
an infinite number of string bits. In this {\it string bit basis},
the Hamiltonian is diagonal.
It is therefore natural to construct the
$N$-string vertex in this basis.
Using this vertex, we construct the corresponding Moyal structure
following the ideas of \cite{Bars,DLMZ}.
We find that the Witten $\ast$-product is equivalent to
a Moyal product and the Moyal structure consists of
infinite copies of 2-dimensional planes with the
same noncommutativity parameter $\th=2$. This is one of the main
result of this paper.
This construction using the string bit basis leads us to
the observation that the equivalence of the Witten $*$-product with
the above-mentioned Moyal product is indeed {\it background
independent}.
% p3
% This equivalence is basis dependent and holds only in
% the string bit basis.
This is another main result of this paper.

The paper is organized as follows. In section 2, we perform the
tensionless limit of the bosonic open string in the
Nappi-Witten type pp-wave background.
In section 3, we construct the $N$-string vertex
in the string bit basis
by solving the overlap conditions.
% p5
It turns out that the Witten $\ast$-product
is equivalent to infinite copies of the Moyal product
with the same noncommutativity $\th = 2$, together with the
midpoint coordinates which are commutative.
In section 5 we point out that the Witten $\ast$-product
for flat background approaches to the same $\th = 2$ algebra
in the high energy limit.
In fact, in the string bit basis,
one can always identify Witten's product with this algebra.
The open string field theory action is constructed in section 6.
%c6 and
Projective operators corresponding to D-branes are
discussed in section 7.
Finally we conclude in section 8.

\section{Tensionless String in pp-wave Background}
\label{worldsheet}

The Nappi-Witten type string model is a  WZW model based on the
extended Heisenberg group $H_n$ \cite{KM}.
The target space $H_n$ is of
dimension $D=2n+2$ and has the metric \eq{pp-metric}.
There is also an NS flux
\be \label{pp-flux}
H_{+(2i+1)(2i+2)} = \mu, \quad i =0, \cdots, n .
\ee

The string action in the conformal gauge takes the form
\bea \label{CGS}
S' = \frac{1}{4\pi \a'}\int d\tau \int_0^{\pi} d\s &[&
\dot{X}^2-{X'}^2-\mu^2 X^2 (\dot{X}^{+2}-{X'}^{+2})
-4(\dot{X}^+\dot{X}^- - {X'}^+ {X'}^-) \nn \\ &&+\;iB_{\mu\nu}
(\dot{X}^{\mu}{X'}^{\nu}-\dot{X}^{\nu}{X'}^{\mu})\;],
\eea
where the ``level'' $k$ has been identified with $\a'^{-1}$.
To take the tensionless limit
\be
\a' \to \infty,
\ee
we rescale the time coordinate
\be
t = \a'\tau .
\ee
Noticing that $B_{\m\n}$ is independent of $\a'$, we obtain
\be \label{Act}
S = \frac{1}{4\pi}\int dt \int_0^{\pi} d\s \left( (\del_t{X})^2 -
(\mu\del_t{X}^+)^2 X^{i2} - 4\del_t{X}^+\del_t{X}^- \right)
\ee
in the limit $\a' \rightarrow \infty$.
As it can be seen clearly from this action,
the string is essentially a combination of
infinitely many independent points labelled by $\s\in [0, \pi]$.
Also note that a new symmetry corresponding
to arbitrary permutation of $\s$ emerges in this limit.

According to (\ref{Act}), the Hamiltonian is
\be \label{Ham}
H = \pi \int_0^{\pi} d\s
\left( P^{i\,2}+\left({\mu P_-\over2}\right)^2 X^{i\,2} - P_+ P_- \right),
\ee
where the conjugate momenta are
\be
P_- ={-1\over \pi} \del_t X^+, \quad P_+ = {-1\over \pi}(\del_t
X^- + \frac{1}{2} \mu^2 X^{i\,2} \del_t X^+), \quad
P^i={1\over2\pi}\del_t X^i.
\ee
The equations of motion are
\bea
&\del_t^2 X^i + (\pi \mu P_-)^2 X^i = 0, \\
&\del_t P_+ = 0, \quad \del_t P_- = 0.
\eea
Due to translational invariance in $X^+, X^-$,
the light-cone momenta $P_+, P_-$ are conserved. It is clear that
each transverse coordinate $X^i(\s)$ at a given point
$\s$ on the string corresponds to a simple harmonic oscillator
\be
a^i(\s) : = \sqrt{\frac{\pi}{ \om_0(\s)}}
\left( P^i(\s) - i\frac{\om_0(\s)}{2 \pi} X^i(\s) \right)
\ee
with angular frequency
\be \label{mu0}
\om_0(\s) = \pi \mu |P_-(\s)|.
\ee
The commutation relations of $a^i(\s)$ are
\be
[a^i(\s), a^j(\s')^\dag] = \d^{ij} \d(\s-\s') .
\ee
The part of the Hamiltonian (\ref{Ham}) relevant to $X^i$ is
\be
H_X = \int_0^{\pi} d\s \; \om_0(\s) \sum_i {a^i}^{\dag}(\s)a^i(\s)
\ee
up to normal ordering. The eigenstates of $H$ are spanned by
\be
\bigotimes_{\sigma\in [0,\pi]}
| n^i(\sigma); P_+(\sigma), P_-(\sigma)\ra\;,
\ee
where $n^i$ is the simple
harmonic oscillator label for $X^i$.

In the next section, we will construct the Witten $*$-product in terms
of this pointwise basis of oscillators. Note that in the tensionless limit,
the string becomes ultra-local and there is no need to impose any
boundary condition on the open string described by the action
\eq{Act}.  The pointwise basis is the appropriate basis for describing
the string configurations in the tensionless limit.
It would be insufficient to use the usual
oscillator basis $a_n$ which is defined in terms of a Fourier basis
$\cos n \s$, since Neumann boundary condition is implied if one use this basis.
Moreover, as we will show,  the pp-wave Hamiltonian is diagonalized
in this basis, it is therefore also an useful choice.
Most importantly, we will show in section \ref{BI}
that written in this basis,
Witten's
$*$-product has a simple universal representation in terms of the
Moyal product, independent of the string background.
Hopefully this will provide some insights into the algebraic
structures of the open string field theory
%c3
at finite $\alpha'$.

We remark that to recover the usual string picture with a finite
$\alpha'$, we
expect that a nontrivial  perturbation ${\cal O}$  can be introduced
to the tensionless theory
\be \label{pert}
S = S_0 +  \lambda {\cal O}, \quad \lambda \sim l_s^{-1}
\ee
that implement the
effects of having a nonzero tension.
The perturbation ${\cal O}$ should introduce
correlation for neighbouring string bits and implement the Neumann
boundary condition. In this paper, we will be contended with the
zeroth order construction.

It is also useful to use a string-bit regularization by
discretizing the
worldsheet into a large number $J$ of bits and write
\be \label{bit1}
a(\s) = \sqrt{\frac{J}{\pi}}\sum_{{\rm c}=0}^{J-1} a_{\rm c} \chi_{\rm c}(\s),
\ee
where $\chi_{\rm c}(\s)$ is a step function with support on the interval
$[\pi {\rm c}/J, \pi({\rm c}+1)/J]$.
In this regularization,
\be \label{bit2}
\int d \s \to \frac{\pi}{J} \sum_{\rm c},
\ee
and string excitations in the transverse directions are characterized by the
oscillators $a^i_c$ obeying
\be \label{bit3}
[a^i_{\rm c},a_{\rm d}^{j\dag}] = \d^{ij} \d_{{\rm cd}},
\quad {\rm c,d} =0, \cdots, J-1.
\ee
We will present our results in the following in terms of $a^i(\s)$,
with the understanding that the discrete basis is employed whenever it is
necessary, e.g. \eq{V3-1}.

Finally we remark that the usefulness of the
pointwise basis $a(\s)$ was also pointed out in
\cite{KO}, where the $N$-string vertex
%c3 for the flat case
were constructed, and it was argued to be a convenient choice
since it  gives additional insights
into the algebraic structure of the open string field theory.

\section{$N$-String Vertex}

Up to an overall normalization,
the $N$-string vertex is determined by the overlap conditions,
\bea
&&\left( X^\m_I (\s) - X^\m_{I-1}(\pi-\s) \right) | V_N \rangle =
0, \label{cc1} \\
&&\left( P^\m_I(\s) + P^\m_{I-1}(\pi-\s) \right) | V_N \rangle = 0,
\quad \s\in[0, \pi/2],
\label{cc2}
\eea
where $I \in {\bf Z}_N$ is the label for different strings, and
$X_I$ and $P_I$ are the string coordinates
and conjugate momenta for the $I$-th string.

% p5
We now assume that the full solution to \eq{cc1} and \eq{cc2} is
of the form
\be \label{VN}
\ket{V_N} =\ket{V_N}_{\pm} \ket{V_N}_X.
\ee
%where the transverse part $\ket{V_N}_X $ is given by
%\be \label{VNt}
%%c5
%\ket{V_N}_X = \hat{U}  \ket{v_N}_X := e^{R_N} |0\ra_X, \quad
%\mbox{with} \quad \hat{U}:=\prod_{I=1}^N U_I
%\ee
%and
It is easy to see that
the light cone part $\ket{V_N}_\pm $ is given by
\be \label{VNl}
|V_N\ra_{\pm} = \prod_{I=1}^N \prod_{\s\in[0,\pi/2]}
\delta(\P_{I}(\s)+\P_{(I-1)}(\pi-\s))
% \ket{P_{\pm 1}, \cdots, P_{\pm N}}
\ket{\P_{1}}\otimes \cdots \otimes \ket{\P_{N}},
\ee
where $\ket{\P_{I}}$ is a (light cone) momentum eigenstate with
eigenvalue $\P_{I}$ for the $I$-th string.
%c4
We recall that the Hamiltonian \eq{Ham} is diagonal in this basis
$\ket{\P_{I}}$.

%c6
We will skip the index $i$ for transverse directions from now on.
The overlap condition \eq{cc1} and \eq{cc2} implies that
\be
%c6 I don't get the i
\frac{1}{\sqrt{\mu |P_{-I}(\s)|}}\ket{V_N}_{\pm}\;\cdot\;
\left(\tilde{X}_I(\s)-\tilde{X}_{I-1}(\pi-\s)\right)
\ket{V_N}_X = 0,
\ee
\be
\sqrt{\mu |P_{-I}(\s)|} \ket{V_N}_{\pm}\;\cdot\;
\left(\tilde{P}_I(\s)-\tilde{P}_{I-1}(\pi-\s)\right)
\ket{V_N}_X = 0,
\ee
where
\be
\tilde{X}_I(\s) := i\left(a_I(\s)-a^{\dag}_I(\s)\right),
\quad \tilde{P}_I(\s) := \frac{1}{2}\left(a_I(\s)+a^{\dag}_I(\s)\right).
\ee
%c6 For later use we also define the conjugate of $\tilde{X}$ as
% \be
% \tilde{P}_I(\s) := \frac{1}{2}\left(a_I(\s)+a^{\dag}(\s)\right).
% \ee
The overlap condition for the transverse part $\ket{V_N}_X$
is thus
%c6
\bea
&& \left(\tilde{X}_I(\s)-\tilde{X}_{I-1}(\pi-\s)\right) \ket{V_N}_X = 0, \\
&& \left(\tilde{P}_I(\s)-\tilde{P}_{I-1}(\pi-\s)\right)
\ket{V_N}_X = 0, \quad \s \in [0,\pi/2].
\eea

The overlap condition relates a point $\s$ on the string
to the point $(\pi-\s)$ on another string.
The only fixed point of the map $\s \rightarrow (\pi-\s)$
is the midpoint $\s = \pi/2$,
and thus the midpoint should be dealt with separately.
The vertex should be decomposable as
% p5
\be \label{vN}
|V_N\ra_X = |V_N^0\ra_X |V'_N\ra_X,
\ee
where $|V_N^0\ra_X$ and $|V'_N\ra_X$ are the vertices
for the midpoint and the rest of the string, respectively.
The overlap condition for the midpoint is simply to
identify $\tilde{X}^i_I(\pi/2)$ for all $I$.
Thus it is easy to find
\bea
|V_N^0\ra_X &=& \int dx |x\ra \otimes \cdots \otimes |x\ra \nn\\
&=&  \int \prod_{I=1}^N \frac{dp_I}{{2\pi}}
\delta\left(\sum_{J=1}^N p_J\right)
|p_1\ra \otimes \cdots \otimes |p_N\ra,
\label{vN0}
\eea
where $|x\ra$ and $|p\ra$ are the eigenstates of
$\tilde{X}(\pi/2)$ and $\tilde{P}(\pi/2)$.
The overlap conditions
%cc \eq{cct1} and \eq{cct2} for the transverse coordinates
for $\s\neq\pi/2$ can be solved by
\be \label{V3-1}
\ket{V'_N}_X = \exp
\left[ - \sum_{i=1}^N
\int_{0\leq\s<\pi/2} d\s
a_I^{\dag}(\s)a_{I-1}^{\dag}(\pi-\s) \right]
|0\ra_X,
\ee
where the vacuum $|0\ra_X$ is defined to be
\be
a_I(\s) \ket{0}_X =0, \quad \mbox{for}\;\;
0\leq\s\leq\pi \;\mbox{and}\; \s\neq\pi/2.
\ee
Here the formula \eq{V3-1}
is interpreted in terms of the string bit picture
\eq{bit1}-\eq{bit3} since it is crucial to treat the
midpoint separately.
We note that,
when acting on $\ket{V_N}_X$, it is
\be \label{ss}
a_I(\s) \simeq -a_{I-1}^{\dag}(\pi-\s) ,
\quad
a_{I-1}(\pi-\s) \simeq -a_I^{\dag}(\s), \quad
\mbox{for $\s \in [0,\pi/2)$}.
\ee

\section{Witten $*$ as Moyal $\ast$} \label{w}

In this section, we will first review how the Moyal
$*$-product is obtained from the Witten $*$-product
%c4
in section \ref{w1}.
In the original work \cite{Bars, DLMZ},
this relation is performed using the harmonic oscillator basis.
Then in section \ref{w2} we will show that a slight variation of the
method can be adapted
for the string bit basis, which is the most natural and convenient
basis in the tensionless limit. A Moyal structure with infinite
copies of noncommutative planes with $\th=2$ is obtained.

\subsection{The Stereotype}\label{w1}

Let us start with the
flat case and with
a single pair of string oscillators.
To interpret Witten's $\ast$-product as the
Moyal product with noncommutativity $\th$ \cite{DLMZ},
we recall that one need to put $|V_3\ra$ in the standard form
\bea
|V_3\ra &=& \frac{2}{3\sqrt{\pi}}\frac{1}{1+\frac{\th^2}{12}}
\exp\left\{\sum_{I=1,2,3} \left[
-\frac{1}{2}\left(\frac{-4+\th^2}{12+\th^2}\right)
(e^{\dag}_I e^{\dag}_I + o^{\dag}_I o^{\dag}_I)\right.\right. \nn \\
&& \left.\left. -\left(\frac{8}{12+\th^2}\right) (e^{\dag}_I
e^{\dag}_{I+1} + o^{\dag}_I o^{\dag}_{I+1})
-\left(\frac{4i\th}{12+\th^2}\right) (e^{\dag}_I o^{\dag}_{I+1} -
o^{\dag}_I e^{\dag}_{I+1}) \right]\right\}|0\ra, \label{stand}
\eea
for some creation and annihilation operators $(e, e^{\dag}, o,
o^{\dag})$ satisfying
\be
[o,o^\dag] =[e,e^\dag] =1, \quad [o,e^\dag]= [e,o^\dag] =0.
\ee
The associated Moyal product is defined on a two dimensional plane
$(y,z)$, where $(y,z)$ label the spectrum of the position operators
$\hat{y}, \hat{z}$
defined by ($[\hat{y}, \hat{z}]=0)$,
\be
\hat{y} := \frac{i}{\sqrt{2}}(e-e^{\dag}), \quad
\hat{z} := \frac{i}{\sqrt{2}}(o-o^{\dag}).
\ee
The corresponding  eigenstate is
\be
\la y,z| = \frac{1}{\sqrt{\pi}}\la 0|
\exp\left(-\frac{1}{2}(y^2+z^2) + i \sqrt{2} (ey+oz) + \frac{1}{2}
(ee+oo) \right).
\ee
The coordinates $y,z$ satisfy
the $\ast$ commutation relation
\be
[y, z]_{\ast} = i\th,
\ee
where the Moyal $\ast$-product is defined through
$|V_3\ra$ according to the correspondence
\be \label{wm-star}
(\la f_1|\otimes\la f_2|\otimes\la y_3, z_3|)|V_3\ra =
(f_1 \ast f_2)(y_3, z_3),
\ee
where
\be
\la f| = \int dy dz f(y,z) \la y,z|.
\ee
The noncommutativity parameter $\th$ is directly related to the
spectrum of the Neumann matrices that appears in the 3-strings vertex.

The above consideration for a single pair
of string oscillators $(e,o)$
can be easily generalized \cite{DLMZ} to the full string vertex.
In the flat case, $|V_3\ra$ in the oscillator basis takes the form
\be
|V_3\ra = \exp\left( -\frac{1}{2}\sum_{I=1}^3
a^{\dag}_I U^{IJ} a^{\dag}_J \right) |0\ra,
\ee
and the following constraints are satisfied
\bea
&U^{IJ} = U^{(I+1)(J+1)}, \label{UU}\\
&M^{12}+M^{21} = 1 - M^{11}, \\
&M^{12}M^{21} = M^{11} (M^{11} -1), \label{UU3}
\eea
where
\be
M^{11} = CU^{11}, \quad M^{12} = CU^{12}, \quad M^{21} = CU^{21}
\ee
and $C$ is the matrix
\be
C_{mn} = (-1)^n \d_{mn}.
\ee
It was shown that the Witten $\ast$-product can be interpreted
as a Moyal $\ast$-product in terms of suitable variables
% p3
\cite{DLMZ}.
The noncommutativity parameter $\th$ is related to the eigenvalues
%c6 change ij to IJ
$\lam^{IJ}$ of the Neumann matrices $M^{IJ}$ by
\be
\lambda^{11} = \frac{-4 + \th^2}{12 + \th^2}, \quad
\lambda^{12}+\lambda^{21} = \frac{16}{12 + \th^2}, \quad
\lambda^{12}-\lambda^{21} = \frac{8\th}{12 + \th^2}.
\label{mu}
\ee
In general, if the 3-string vertex satisfies the properties
(\ref{UU})--(\ref{UU3}), we can hope that Witten's product
can be written as a Moyal product.

\subsection{Moyal Product for pp-wave} \label{w2}

To put $|V_3\ra_X$ of (\ref{V3-1}) in the standard
form (\ref{stand}), we define a new basis of oscillators
% p3
$\{ e(\s), o(\s)\}$
by
\be
a^{\dag}(\s) = \frac{1}{\sqrt{2}}(e^{\dag}(\s) + i o^{\dag}(\s)),
\quad
a^{\dag}(\pi-\s) = \frac{1}{\sqrt{2}}(e^{\dag}(\s) - i o^{\dag}(\s)),
\quad \s\in[0, \pi/2).
\ee
The oscillators $e,o$ satisfy the commutation relations
\be
[o(\s),o^\dag(\s')] =[e(\s),e^\dag(\s')] = \d(\s-\s'),
\quad [o(\s),e^\dag(\s')]= [e(\s),o^\dag(\s')] =0.
\ee
Then one introduces the infinite copies of 2-dimensional planes whose
coordinates $y(\s), z(\s)$ are the eigenvalues of the position
operators ($[\hat{y}(\s), \hat{z}(\s')] =0$),
\bea \label{yz}
&&\hat{y}(\s)=
\frac{i}{\sqrt{2}}(e(\s)-e^{\dag}(\s))=\frac{1}{2}(\tilde{X}
(\s)+\tilde{X}(\pi-\s)),
\\
&&\hat{z}(\s)=
\frac{i}{\sqrt{2}}(o(\s)-o^{\dag}(\s))=
- \tilde{P}(\s) +\tilde{P}(\pi-\s),
\quad \mbox{for $\s \in [0, \pi/2)$}.
\eea
Then one sees that the Witten $\ast$-product defined by
%c6 $|v_3\ra_X$
$|V_3\ra_X$
is equivalent to the
Moyal $\ast$-product
\be \label{cr}
[ y(\s), z(\s') ]_{\ast} = 2i \delta(\s - \s'), \quad \s, \s' \in [0, \pi/2)
\ee
with the same noncommutativity parameter
$\th = 2$.
Note that the algebra (\ref{cr}) is not defined
for $\s = \pi/2$ because $z(\pi/2) = 0$, that is, the midpoint
$y(\pi/2)=\sqrt{2}X(\pi/2)$ is a single commutative coordinate.
This is natural because there is no distinction
between left and right at the midpoint.
We shall denote the midpoint coordinate by $x \equiv y(\pi/2)$.
% p3
%We note that in the basis adopted in \cite{DLMZ},
%it is the momentum carried by the left half
%of the string which behave as a commuting coordinate.

To display the Moyal product more explicitly,
let us define the string wave function $\Psi(x,y,z,P_+,P_-)$
using the basis
\be \label{basis1}
% p5 {}_U
\la x,y,z,\P|
:=
\la x| \la y,z| \la P_+, P_-|
% U^{-1}
.
\ee
That is,
\be
\la \Psi| = \int dx Dy Dz D \P \;\; \Psi(x,y,z,\P)
% p5 {}_U
\la x,y,z,\P|,
\ee
where
\be
D\P = \prod_{\s\in[0,\pi]}dP_+(\s) dP_-(\s).
\ee
% p5
%One can show that
%\bea
%&{}_U \la x,y,z,\P|\left(X(\s)+X(\pi-\s)\right) =
%2y(\s){}_U \la x,y,z,\P|, \\
%&{}_U \la x,y,z,\P|\left(P(\s)-P(\pi-\s)\right) =
%-z(\s){}_U \la x,y,z,\P|
%\eea
%for $\s < \pi/2$.
%%&{}_U \la x,y,z,\P|P^{\pm}(\s) =
%%P^{\pm}(\s){}_U \la x,y,z,\P|.

The full Moyal $*$-product is defined through
the Witten $*$-product by
\be
(\Psi_1\ast_M\Psi_2)(x,y,z,\P) =
\left(\la\Psi_1|\otimes\la\Psi_2|\otimes
% p5 {}_U
\la x,y,z,\P|\right)|V_3\ra.
\ee
It turns out to be
\bea
(f\ast_M g)(x,y,z,\P) &=&
\int D\P_1 D\P_2 \;f(x,y,z,\P_1)\ast g(x,y,z,\P_2)
\d(\P(\s)+\P_2(\pi-\s)) \nn \\
&& \d(\P_2(\s)+\P_1(\pi-\s)) \d(\P_1(\s)+\P(\pi-\s)) \label{Moyal}
\eea
where $\ast$ for the 2-dimensional planes is defined by (\ref{cr}):
\be
e^{i \int_{0\leq\s<\pi/2} d\s
(\del_{y_1} \del_{z_2}-\del_{z_1} \del_{y_2})}
%c4 |_{y_1=y_2=y_3,\\ z_1=z_2=z_3}.
 |_{y_1=y_2=y ,\\ z_1=z_2=z }.
\ee

We note that the noncommutative algebra (\ref{cr})
is the same as the large $\kappa$ limit of the
flat background case \cite{DLMZ} (see also \eq{flat} below).
This is not surprising because the tensionless limit is
a high energy limit. We will have more comments  on this in the next section.
We also note that this algebra possesses the permutation symmetry $S_{\infty}$
that allows one to exchange
any two points $\s$ and $\s'$ on the string
for $\s, \s' < \pi/2$, with their mirror images (with respect to the
midpoint of the string)
$(\pi-\s)$ and $(\pi-\s')$ also swapped at the same time.
For this to be the symmetry of the string field theory,
we need to check that the BRST operator is invariant.
The BRST operator in Segal's gauge is
%c4 given below in (\ref{L0}),
constructed below from (\ref{L0}),
and this is indeed the case.

%c3
% Finally we note that one may also choose a different set of variables.
% For a complete orthonormal basis of functions
% $\{f_n\}$ on $[0,\pi/2)$,
% we can define
% \be
% y_n = \int_0^{\pi/2} d\s y(\s) f_n(\s), \quad
% z_n = \int_0^{\pi/2} d\s z(\s) f_n(\s),
% \ee
% and then
% \be
% [y_m, z_n]_{\ast} = 2i \delta_{mn}, \quad [y_m, y_n]_{\ast} =
% [z_m, z_n]_{\ast} = 0.
% \ee
% For comparison with the flat background results \cite{DLMZ},
% one may choose the bases $\{\cos(2n\s)\}$ and $\{\cos((2n-1)\s)\}$
% for $y$ and $z$, respectively.

\section{Background Independence of Star Product} \label{BI}

In this section, we will show that the Witten $*$-product,
when expressed in terms of the string bit basis, can
always be written in terms of the universal Moyal structure
with $\th = 2$.

How is the algebra (\ref{Moyal})
related to the algebra for the flat background?
The Neumann matrices are diagonalized
for the flat background in \cite{RSZ}
(apart from the zero mode)
and the corresponding Moyal product has \cite{DLMZ}
\be \label{flat}
\th = 2 \tanh(\pi \kappa/4),
\ee
%c4
where $\kappa\in[0, \infty)$
is the index for eigenvectors.
The eigenvectors $v_n(\kappa)$ for the flat background
can be obtained from the generating function
\be
f_{\kappa}(z) = \sum_{n=1}^{\infty}
\frac{v_n(\kappa)}{\sqrt{n}} z^n \propto
(1 - \exp(-\kappa \tan^{-1} z)).
\ee
When we increase the parameter $\kappa$,
the eigenvector $v(\kappa)$ tends to
have more contribution from larger oscillation modes.
The peak of $v_n(\kappa)$ is at around $n\sim \log\kappa$
for large $\kappa$.
As a result, roughly speaking,
in the high energy limit
we should consider modes with large $\kappa$,
and for $\kappa \rightarrow \infty$,
the parameter $\th$ has a finite limit at $\th = 2$.
Therefore we find that the Witten product corresponds
to the Moyal product with $\th = 2$ for both pp-wave
and flat backgrounds in the high energy limit.
This suggests that the Witten $*$-product can
always be written in terms of this universal Moyal structure
for any background in the tensionless limit.
As we shall see, it is even more general than this.

It is useful to recall that the great simplification in using
the string bit basis is that it provides a very clear
exposition of the algebraic structure of the Witten product.
Moreover it also
diagonalizes the Hamiltonian in the tensionless limit.
Therefore it is not just natural, but also
convenient to use this basis to construct the string
vertices and to represent the Witten product in the high energy
limit. When $\a'$ is finite, the string bit basis no longer diagonalizes
the Hamiltonian. However its algebraic advantages remain.
%c3
In fact, for any background,
the Witten $*$-product written in the string bit basis
can always be presented in terms of the Moyal
product with the universal noncommutativity parameter $\th=2$.

It is easy to see this.
Note that for any background, one can simply use
the string bit basis of creation and annihilation operators
$a(\s), a^\dag(\s)$ defined by
\be \label{XPtilde}
%c6
% p6
X(\s) = i (a(\s) - a^{\dag}(\s)), \quad
P(\s) = \frac{1}{2}(a(\s) + a^{\dag}(\s)),
\ee
and define the state $|0\ra$ by $a(\s)|0\ra = 0$.
Immediately we find that the N-string vertices
given by the same expressions
(\ref{vN}), (\ref{vN0}) and (\ref{V3-1}) for
%c6
$|V_N\ra$
satisfy all the overlap equations.
We stress that this holds for any $\alpha'$.

In the literature, the $N$-string vertices are often
constructed in terms of the perturbative vacuum and
the creation, annihilation operators for the perturbative spectrum,
which are useful for a perturbative description.
However, as a nonperturbative formulation,
the string field theory is background independent.
It is not always necessary to write down everything
in accordance with the perturbative spectrum.
What we just showed is that, independent of the string background,
one  can always represent
the Witten product as the Moyal product
with $\th = 2$ in the string bit basis;
and that in the tensionless
limit, this basis also diagonalizes the Hamiltonian and so should be a
useful one.
%c3
We expect that this background independent representation of the Witten
$*$-product as a
universal Moyal product  should provide insights
into a better understanding of
the formal aspects of Witten's open string field theory.

%c3
Before we move on,
we remark that in a generic background,
the permutation symmetry $S_{\infty}$ is
preserved only in the high energy (tensionless) limit and is always broken
in the low energy by the kinetic term.
It will be interesting to see if there is any background in which
this global symmetry is unbroken and
the Moyal product takes the simple form even for finite $\a'$.

\section{String Field Theory Action}

We focus on the matter part of the string field theory action,
which has the following kinetic term in the Siegel gauge
($b_0 \Psi = 0$)
\be
\frac{1}{2\a'} \int \Psi (L_0-1) \Psi\;.
\ee
In the pp-wave background with $\a'\rightarrow\infty$,
$L_0 = \a' H$ (\ref{Ham})
is the Virasoro operator with respect to $\tau$
% p5 .
%In the representation $\Psi(x,y,z,\P)$
%we have to replace $L_0$ by
\be \label{L0}
% p5 U^{-1}L_0 U
%= \pi \a' \int_0^{\pi} d\s \left(
%\tilde{P}^2 + (\frac{\mu P_-}{2})^2 \tilde{X}^2 -P_+ P_- \right)\equiv
%\a' L'_0.
L_0 = \pi \a' \int_0^{\pi} d\s \left(
P^2 + (\frac{\mu P_-}{2})^2 X^2 -P_+ P_- \right) .
\ee
One can rewrite everything in terms of $x,y,z,\P$ by
defining the derivatives of $y(\s), z(\s)$
%cc factor corrected
\be
\frac{\del}{\del y(\s)} =
i (\tilde{P}(\s)+\tilde{P}(\pi-\s)), \quad
\frac{\del}{\del z(\s)} =
\frac{i}{2}(\tilde{X}(\s)-\tilde{X}(\pi-\s))
\ee
and use linear combinations of $y$, $z$ and
their derivatives to express
% p5
$P$ and $X$
in (\ref{L0}).
% p6
In the large $\a'$ limit we can replace $(L_0-1)$ by $L_0$
for the kinetic term,
and the factor of $\alpha'$ in $L_0$ will cancel
the $\alpha'$ in the denominator of the kinetic term in (\ref{ActSFT}).

Including the cubic term,
the full string field theory action in the tensionless limit is
\be \label{ActSFT}
S={-1\over g^2} \left({1\over 2}\int \Psi
% p6
H
\Psi
+{1 \over 3}\int \Psi *_M \Psi *_M \Psi\right),
\ee
where $g$ is the open string coupling, and
\be
H = \pi \int_0^{\pi/2} d\s \left[
-\frac{1}{2}\mu|P_-(\s)|(\del_y^2 + \del_z^2 - y^2 - z^2)
+ P_+(\s)P_-(\s) + P_+(\pi-\s)P_-(\pi-\s) \right].
\ee
% p6
%In order to keep $L_0$ finite in the limit $\a' \rightarrow \infty$,
%we scale $P_-(\s)$ simultaneously so that
%$p(\s) := \a' P_-(\s)$ is finite in the limit.

% p5
%Denoting
%\be
%s^2(\s) = P^2_-(\s) + P^2_-(\pi-\s), \quad
%t^2(\s) = P^2_-(\s) - P^2_-(\pi-\s),
%\ee
%we find
%\be
%L'_0 = \pi\int_0^{\pi/2} d\s \left[
%-\frac{1}{2}(D_y^2 + D_z^2 - z^2 - \kappa^2 y^2)
%+ P_+(\s)P_-(\s) + P_+(\pi-\s)P_-(\pi-\s) \right],
%\ee
%where
%\bea
%&&D_y = \frac{\del}{\del y(\s)}, \quad
%D_z = \frac{\mu s}{\sqrt{2}}\left( \frac{\del}{\del z(\s)} +
%i\frac{t^2}{s^2} y(\s) \right), \\
%&&\kappa = \frac{\sqrt{2}\mu}{s(\s)} P_-(\s)P_-(\pi-\s).
%\eea
%%The Bogoliubov transformation (\ref{Bogo}) by $U$
%%corresponds to turning on the nonvanishing curvature
%There is a nonvanishing background field strength
%\be
%[D_y, D_z] = i\frac{\mu}{\sqrt{2}}\frac{t^2}{s}.
%\ee

\section{Projective Operators and D-branes}

% ppp I put the discussion on the ratio of tensions to the end

For the flat background,
it was conjectured that D-branes are projective states \cite{RSZ}
in the vacuum string field theory (VSFT) \cite{VSFT}.
It has been checked that, in addition to giving
the correct ratio of brane tensions \cite{RSZ,RSZ1},
the tachyon state of open strings ending on D-branes
are reproduced \cite{HK,RSZ2,Okawa} from the VSFT.
In this section we will follow the assumptions and formulations of
the vacuum string field theory (VSFT) \cite{VSFT} in flat space
and consider projections as candidates of D-branes
in the pp-wave background for the tensionless limit.

The projections are usually constructed in the oscillator basis,
but we will see that it is easier to consider them
as projective functions of the Moyal variables.
Due to the simplicity of our Moyal product,
our job here is much simpler than the flat case.

Since the midpoint is a commutative coordinate as noted before,
it is more convenient to separate it from the others
and define the variables
%c3 p changed to P below
\bea
Y(\s) &=& y(\s) -
\frac{1}{\sqrt{2 P_-(\pi/2)}}\left(\sqrt{P_-(\s)}
+\sqrt{P_-(\pi-\s)}\right) x, \\
Z(\s) &=& z(\s), \\
x &=& \sqrt{\mu P_-(\pi/2)}X(\pi/2),
\eea
so that
\be
\label{nc1}
[Y(\s), Z(\s)]_* = 2i \delta(\s-\s'), \quad
[x, Y(\s)]_* = [x, Z(\s)]_* = 0,
\ee
and more importantly,
$Y(\s), Z(\s)$ are invariant under
% p3
a
shift of $X(\s)$.
A string state $\Psi$ can be viewed as a function of
% p3 the noncommutative pairs
$Y(\s), Z(\s), \P(\s)$
and the midpoint coordinate $x$.
Based on the relation (\ref{nc1}),
we consider projective states which take the  factorized form:
%c3 are factorized as
\be \label{factor}
\Psi(x,Y,Z,\P) = \Phi(\P)\Phi'(Y,Z)\Phi''(x)\;,
\ee
where each factor is a projective function.

First we consider the factor $\Phi(\P)$,
which can be further decomposed as
\be
\Phi=h\cdot g,
\ee
where $h(\P(\pi/2))$ depends only
on $\P(\pi/2)$ and $g(\P)$
depends on $\P(\s)$ for $\s\neq\pi/2$.
For $g$ to be a projection, we need
\bea \label{gi}
g(\P_3) &=& \int D\P_1 D\P_2 \;g(\P_1)g(\P_2)
\d(\P_1(\s)+P_3(\pi-\s)) \nn\\
&&\d(\P_2(\s)+\P_1(\pi-\s))\d(\P_3(\s)+\P_2(\pi-\s)).
\eea
The simplest possibility is for $g$ to be a constant,
independent of $\P$.
A few less trivial cases are:
\bea
g(\P)&\propto&\d(\P(\pi/2))
\prod_{\s<\pi/2}\d(\P(\s)+\P(\pi-\s)), \\
g(\P)&\propto&\d(\P_L)\d(\P_R), \label{gP} \\
g(\P)&\propto&\d(P_+(\s)-P_-(\s)).
\eea
%It is also possible to choose the $P_+$ and $P_-$
%dependence separately to be either of the two types above.
The condition for $h$ analogous to (\ref{gi}) is
\be
h(\P_3)=\int d\P_1 d\P_2 h(\P_1)h(\P_2)
\d(\P_1+\P_2)\d(\P_2+\P_3)\d(\P_3+\P_1).
\ee
It follows that
\be \label{hi}
h(\P(\pi/2)) \propto \d(\P(\pi/2)).
\ee
This implies that the our assumptions
about factorization (\ref{factor})
works only for lumps stretched in the $X^-$ direction.
%and the difference between D$p$-brane and D$(p+1)$-brane
%must lie in the transverse directions.

%c3 Now we consider
For $\Phi'(Y,Z)$, the simplest choice
is to take it to be the noncommutative GMS solitons \cite{GMS}.
For example, $\Phi'(Y,Z)$ can be
%c3 given by
taken as
\be \label{00}
{\cal N} e^{-{1\over2}\int d\s(Y^2(\s)+Z^2(\s))}\equiv
|0\rra \lla 0|\;,
\ee
where ${\cal N}$ is a normalization constant
depending on spectral measure.
%c3 for example,
In the discrete string bit basis
(\ref{bit1}), ${\cal N}=2^{J\over \pi}$.
In the flat case, $|0\rra \lla 0|$
corresponds to the ground state $|0\ra$ of the  perturbative string.
This is
%c4 also
consistent with the result obtained in the
oscillator basis \cite{RSZ}.
There, $T=M=0$ for $\th=2$, so that the sliver state
is proportional to $|0\ra$.
% ppp
%Note also that $T={1\over M}=\infty$ for
%$\th=-2$ which gives the non-normalizable state.
All rank-one projector states such as
$|n \rra\lla n|$ can be obtained from (\ref{00})
by unitary transformations,
and be mapped into the oscillator basis
with respect to $|V_3\ra$.
These states are new projectors of
vacuum string field theory and
are different from the sliver state,
more details on this point for the flat background
can be found in \cite{CL}.

Finally we consider $\Phi''(x)$.
Any state for the midpoint coordinate can be written in the form
\be
\int dx f(x) |x\ra.
\ee
It is a projection if $f^2(x) = f(x)$.
A projection for a lump extending from $x_1$ to $x_2$
is of the form
\be
\label{midpointproj}
\Phi''(x) = \Th(x-x_{1}) - \Th(x-x_{2}),
\ee
for $x_1<x_2$.
$\Th(x)$ is the step function which equals one or zero
depending on whether $x$ is positive or negative.

At this point it is not clear whether
any of these projective states considered above
represents D-branes.
A nontrivial test for the sliver state
to be a D-brane in the context of
vacuum string field theory in flat background \cite{RSZ}
is the decent relation of the D-brane tension, that is
\be
\label{descent}
{T_p \over 2\pi \sqrt{\a^{'}}T_{p+1}}=1 .
\ee
Another more difficult test is to look for open string states
in the background of D-branes \cite{HK,RSZ2,Okawa}.
The latter requires a lot more work,
and we leave it for the future.

Let us first review the sliver states in the flat background.
In the flat case, the longitudinal direction of a D-brane is constructed
with respect to the zero momentum sector of the 3-string vertex
in the momentum basis \cite{RSZ}, denoted as $|V_3\ra_{p_0=0}$.
%The commutation relation
%\be
%[x_0,p_0]=i\;,
%\ee
The D-brane's longitudinal directions are thus infinite
due to the fixed zero momentum.
On the other hand, the transverse direction
is constructed with respect to the 3-string
vertex in the oscillator basis, which is denoted as $|V_3\ra_o$,
and is related to
%ff
$|V_3\ra_{p_0}$ by
\be
x_0={i\over 2 \sqrt{b}}(a_0-a^{\dag}_0)\;, \quad
p_0=\sqrt{b}(a_0+a^{\dag}_0)\;,
\ee
where $b$ represents the thickness of the D-brane which
breaks translational invariance. Although the 3-string
vertex in both bases are mathematically equivalent, the
corresponding sliver states are not.
%because of the zero momentum
%condition which is due to the translational invariance in the
%corresponding direction.
An important result \cite{RSZ} is that the ratio (\ref{descent})
is independent of $b$.
%although there is no clear understanding why
%this should be the case.

In the Moyal basis of \cite{DLMZ}, $|V_3\ra_{p_0=0}$ and $|V_3\ra_o$
are mathematically of the same form (\ref{stand}) in the
continuous spectrum
\footnote{
% p3
The discrete spectrum is not fully explored yet, and it contains
%c4 the
information about the thickness.
%c4 dependence.
} \cite{Belov}. The difference between
the sliver states for D$p$-brane and D$(p+1)$-brane
corresponds to different choices of the
representation for the commutative variables with
$\th(\kappa=0)=0$.
The one for $|V_3\ra_{p_0=0}$ is $P_L-P_R$
% p5
(or equivalently $P_L$, since $p_0=0$) and the one for $|V_3\ra_o$ is the
midpoint coordinate $X({\pi\over2})$.
In \cite{MT}, singularities in the sliver
states describing D-branes in the flat background
was identified \footnote{ It was shown that the
singularities of the silver states are resolved when a
constant $B$-field is turned on \cite{loriano}.
This is related to the noncommutative spacetime
algebra for the zero mode as discussed in \cite{CL0}
and the second of \cite{loriano}.}.
For the transverse directions, the sliver
state satisfies
\be \label{trans}
X(\pi/2)|\Psi\ra = 0.
\ee
For the longitudinal directions
the sliver state is invariant under the variation
\be \label{long}
\delta X(\s) = \left\{
\begin{array}{ll}
\lambda & \mbox{for} \;\; 0\leq \s<\pi/2, \\
-\lambda & \mbox{for} \;\; \pi/2<\s\leq \pi .
\end{array}\right.
\ee
In terms of the momentum, it is
\be \label{PLR}
(P_L - P_R)|\Psi\ra = 0,
\ee
where
\be
P_L \equiv \int_0^{\pi/2}P(\s), \quad
P_R \equiv \int_{\pi/2}^{\pi}P(\s).
\ee

%c3
Back to the pp-wave case,
we will assume that these properties still hold
for the pp-wave background and use them as our major
% p3 guideline
guide
to identify the D-brane projections.
We also assume that the matter part of the string wave function
corresponding to a D$p$-brane
extending along $X_+$, $X_-$, $X_2$, $\cdots$, $X_p$ directions
is of the form
\be \label{Psip}
\Psi_p = \Phi(\P)
\prod_{i=2}^{p}\psi_1(x_i, Y_i, Z_i)
\prod_{i=p+1}^{D}\psi_2(x_i, Y_i, Z_i),
\ee
and that each factor $\psi_i$ ($i=1,2$) factorized as
\be
\psi_i = \Phi'_i(Y, Z)\Phi''_i(x).
\ee

Let us comment on each of the factors in \eq{Psip}.
As we mentioned above, the ansatz only works
for D-branes extended in the light cone directions.
One can check that the projection (\ref{gP})
satisfies this condition (\ref{PLR}).
The factor
%c3 I take out the ``prime'' from \Phi below
$\Phi(\P)$ is thus given by
%c3 (\ref{PLR}) and (\ref{hi}).
(\ref{gP}) and (\ref{hi}).

%c3 change ``other'' to ``the''
For the longitudinal directions,
$x$ should be allowed to end anywhere,
so we should have $\Phi''_1 = 1$
($x_1\rightarrow -\infty, x_2\rightarrow \infty$)
according to (\ref{midpointproj}).
Note that the coordinates $Y$, $Z$ are both
invariant under the variation (\ref{long}),
so the projector $\Phi'_1(Y,Z)$
written in terms of $Y$, $Z$,
such as (\ref{00}), automatically satisfies
the desired criterion.

For the transverse directions,
$x$ is restricted to the place
where the brane is localized.
We take
\be
\Phi''_2(x) = \Th(x+b/2)-\Th(x-b/2)
\ee
% p3
for a D-brane with finite thickness $b$.
In terms of dimensionless quantities,
the condition (\ref{trans}) should be replaced by
\be \label{cpi2}
\frac{X(\pi/2)}{\sqrt{\a'}}|\psi_2\ra = 0.
\ee
We will comment on this condition later.
%c5
% But since we take $\a' \rightarrow \infty$,
% this condition is always satisfied.

%c3 move here
The ratio of D-brane tensions
\be
R := \frac{S_p}{S_{p+1}} =
\frac{T_p V_p}{T_{p+1} V_{p+1}} =
\frac{\la\Psi_p|\Psi_p\ra}{\la\Psi_{p+1}|\Psi_{p+1}\ra}
\ee
is then given by
\be
R= \frac{\la\psi_2|\psi_2\ra}{\la\psi_1|\psi_1\ra}
= \left. \frac{\la\Phi''_2(x)|\Phi''_2(x)\ra}{\la\Phi''_1(x)|\Phi''_1(x)\ra}
\right|_{\P(\pi/2)=0}.
\ee
where we set $\P(\pi/2)=0$ because of (\ref{hi}).
% p3
%{\it    I don't understand how to get the denominator from  eqn (110)
%above. I thought we said it is $\Phi''(x) =1$ in the paragraph
%bove eqn (107). }
%
%Let the D$p$-brane to have a thickness of $b$.
The ratio $R$ is thus
\be
R = \left.\frac{b}{\int dx}\right|_{\P(\pi/2)=0} =
\left.\frac{b}{\int \sqrt{\mu P_-(\pi/2)} dX(\pi/2)}
\right|_{\P(\pi/2)=0} =
\left.\frac{b}{\sqrt{\mu P_-(\pi/2)} L}\right|_{\P(\pi/2)=0},
\ee
where $L$ is the size of the transverse direction,
and so it equals $V_{p+1}/V_p$.
Setting $P_-(\pi/2)=0$ in $R$, we find
\be
\frac{T_p}{T_{p+1}}=\frac{S_p V_{p+1}}{S_{p+1} V_p}
=\frac{b}{\sqrt{\mu P_-(\pi/2)}}=\infty
\ee
for any finite $b$.
This is indeed what we expect since
\be
\label{tension} \frac{T_p}{T_{p+1}} = 2\pi\sqrt{\a'} = \infty
\ee
in the tensionless limit.

In the zero tension limit the string length
becomes infinity, it makes any finite thickness
of the D$p$-brane negligible so that the ratio
(\ref{tension}) is not sensitive to the finite $b$ value.
But rigorously speaking,
the transverse directions of the D-brane states
are not well-defined, because the finiteness of $b$
implies that the extent of $X(\pi/2)$ is
% p3
$b/\sqrt{\mu P_-(\pi/2)} = \infty$ for $P_-(\pi/2) = 0$.
%c5
To satisfy \eq{cpi2}, we have to take
$\alpha'\rightarrow \infty$ and
$P_-(\pi/2) \rightarrow 0$ with
$\a' P_-(\pi/2) \rightarrow \infty$.
That is, a double scaling limit is involved.
We expect this singularity to be
% p5 resolved
renormalized
when we include $1/\a'$ corrections to the calculation.

\section{Conclusion}

% ppp paragraphs reshuffled.

In this paper, we studied and constructed the
open bosonic string field theory
in pp-wave background in the high energy limit.
Following the recent work \cite{Bars,DLMZ},
we showed that the Witten $*$-product
for open string in the pp-wave background also admits
a presentation in terms of the
Moyal product.
We find that the Moyal product has a universal noncommutativity
of $\th=2$ common to all string modes
in terms of the string bit basis.

%c3 paragraph moved here and modifications
We have also obtained the string field theory action in the
tensionless limit.
We plan to carry out a more detailed analysis of its dynamical aspects
in the future. It would also be interesting to extend our results
to the superstring case, which will be more relevant to the new pp/SYM
duality \cite{BMN}.

The simplicity of the Moyal algebra makes it easy to
find projective operators.
We have proposed to identify some of
these projectors with D-branes
in the VSFT.
The ratio of D-brane tensions can be easily calculated,
but in the zero tension limit $T_{p+1}/T_p$ is just zero.
It will be useful if a perturbative method
based on $\a'^{-1/2}$ expansion can be formulated.
Then it may be possible to
establish the more nontrivial result
$T_{p+1}/T_p = 0+(2\pi\a'^{1/2})^{-1}+\cdots$.
A first step towards
this goal is to identify the perturbation in \eq{pert}. The
perturbation will have to break the permutation symmetry $S_{\infty}$.

%c3
We remark that
this point of view of relating the finite tension string theory with the
tensionless theory has the advantage that the the unperturbed
tensionless theory has a much richer symmetry structure, as we
demonstrated in this paper.
This could be a convenient and powerful starting  point for a
background independent formulation of string theory.
Note that the degrees of freedom (the string
bits) employed in the tensionless theory look very different from the
ordinary degrees of freedom in a string theory with finite tension.
However this is perfectly acceptable. It
may be useful to compare the situation with QCD. At zero coupling, the
natural degrees of freedom are the free quarks and gluons. At nonzero
coupling, the physical objects are  colourless such as baryon. It may
be more than an analogy to think of a finite tension string as a bound
state of string bits.

% p4
A distinctive feature of the pp-wave background
is that, as noted in \cite{Miao},
for weak couplings the tensionless string
will not turn into a black hole as in the flat background,
but will turn into giant gravitons.
It would be interesting to see this more explicitly in the SFT.
One may also wonder whether there is a description of the theory
in which the giant gravitons are the fundamental objects.
% p4
On the other hand,
the commutative coordinate in the 3-string vertex we constructed is
convenient for the Dirichlet boundary condition
for the transverse coordinates.
This may suggest another description
in terms of D-instantons in the zero tension limit of the pp-wave SFT.
%It deserves further study to make a connection between
%our string bit formulation of the SFT
%and the matrix model by using the D-instanton like objects.

Very recently the light cone superstring field theory
in pp-wave background is considered in \cite{SV}.
Witten's open string field theory is covariant and the
interaction of strings is characterized by the midpoint.
This is in contrast with the situation in
the light cone string field theory
where interaction is described in terms of
the splitting and jointing of strings.
Another major difference is that
the product in light cone string field theory
cannot be presented as the Moyal product.
Nevertheless it  would be interesting to compare these
results in the high energy limit.

It has been conjectured \cite{Sund} that in the high energy limit
string theory in the $AdS$ background
has infinite higher spin gauge symmetries.
Their holographic duals are higher spin conserved currents
in the Yang-Mills theory.
Such a gauge theory in $AdS$
has already been constructed in \cite{FV}.
It would be interesting to see whether it corresponds to
% p3 (part of)
the string theory.
Incidentally,
in another limit with $N\rightarrow\infty$ and $g^2 N$ fixed,
infinite higher spin gauge symmetries are also expected
in $AdS$ space \cite{JHS60}.
%c4
See also \cite{sezgin}.
In both cases, something like the Higgs mechanism
should occur to break the higher spin gauge symmetries
when the coupling and energies are finite.
We hope that this work will help
%c4
understanding these important issues about
the short-distance physics of string theory.

\section*{Acknowledgment}

The authors thank Miao Li for helpful discussions.
FLL wants to thank Bin Chen for discussions and the
hospitality of KIAS during his visit
%c5 in the period of finishing the draft.
when the draft is finishing. CSC acknowledge Nuffield Foundation for a grant
NUF-NAL/00445/G. The work of PMH and FLL are supported in part by the National
Science Council, Taiwan, R.O.C. the Center for Theoretical Physics at National
Taiwan University, and the CosPA project of the Ministry of Education, Taiwan.

%\baselineskip 22pt
% \baselinestretch 1.2

\end{document}